\documentstyle[12pt]{article}

\catcode`\@=11
\long\def\@makefntext#1{ 
\protect\noindent \hbox to 3.2pt {\hskip-.9pt
$^{{\ninerm\@thefnmark}}$\hfil}#1\hfill} 

\def\thefootnote{\fnsymbol{footnote}}
 \def\@makefnmark{\hbox to 0pt{$^{\@thefnmark}$\hss}}  

\def\ps@myheadings{\let\@mkboth\@gobbletwo
\def\@oddhead{\hbox{} 
\rightmark\hfil\ninerm\thepage}
\def\@oddfoot{}\def\@evenhead{\ninerm\thepage\hfil 
\leftmark\hbox{}}\def\@evenfoot{}
\def\sectionmark##1{}\def\subsectionmark##1{}}

\begin{document}

\newcommand{\symbolfootnote}{\renewcommand{\thefootnote}
	{\fnsymbol{footnote}}}
\renewcommand{\thefootnote}{\fnsymbol{footnote}}
\newcommand{\alphfootnote}
	{\setcounter{footnote}{0}
	 \renewcommand{\thefootnote}{\sevenrm\alph{footnote}}}

\newcounter{sectionc}\newcounter{subsectionc}\newcounter{subsubsectionc}
\renewcommand{\section}[1] {\vspace{0.6cm}\addtocounter{sectionc}{1}
\setcounter{subsectionc}{0}\setcounter{subsubsectionc}{0}\noindent
	{\bf\thesectionc. #1}\par\vspace{0.4cm}}
\renewcommand{\subsection}[1] {\vspace{0.6cm}\addtocounter{subsectionc}{1}
	\setcounter{subsubsectionc}{0}\noindent
	{\it\thesectionc.\thesubsectionc. #1}\par\vspace{0.4cm}}
\renewcommand{\subsubsection}[1] {\vspace{0.6cm}
\addtocounter{subsubsectionc}{1}
	\noindent {\rm\thesectionc.\thesubsectionc.\thesubsubsectionc.
	#1}\par\vspace{0.4cm}}
\newcommand{\nonumsection}[1] {\vspace{0.6cm}\noindent{\bf #1}
	\par\vspace{0.4cm}}

\newcounter{appendixc}
\newcounter{subappendixc}[appendixc]
\newcounter{subsubappendixc}[subappendixc]
\renewcommand{\thesubappendixc}{\Alph{appendixc}.\arabic{subappendixc}}
\renewcommand{\thesubsubappendixc}
	{\Alph{appendixc}.\arabic{subappendixc}.\arabic{subsubappendixc}}

\renewcommand{\appendix}[1] {\vspace{0.6cm}
        \refstepcounter{appendixc}
        \setcounter{figure}{0}
        \setcounter{table}{0}
        \setcounter{equation}{0}
        \renewcommand{\thefigure}{\Alph{appendixc}.\arabic{figure}}
        \renewcommand{\thetable}{\Alph{appendixc}.\arabic{table}}
        \renewcommand{\theappendixc}{\Alph{appendixc}}
        \renewcommand{\theequation}{\Alph{appendixc}.\arabic{equation}}
        \noindent{\bf Appendix \theappendixc #1}\par\vspace{0.4cm}}
\newcommand{\subappendix}[1] {\vspace{0.6cm}
        \refstepcounter{subappendixc}
        \noindent{\bf Appendix \thesubappendixc. #1}\par\vspace{0.4cm}}
\newcommand{\subsubappendix}[1] {\vspace{0.6cm}
        \refstepcounter{subsubappendixc}
        \noindent{\it Appendix \thesubsubappendixc. #1}
	\par\vspace{0.4cm}}

\def\abstracts#1{{
	\centering{\begin{minipage}{30pc}\tenrm\baselineskip=12pt\noindent
	\centerline{\tenrm ABSTRACT}\vspace{0.3cm}
	\parindent=0pt #1
	\end{minipage} }\par}}

\newcommand{\bibit}{\it}
\newcommand{\bibbf}{\bf}
\renewenvironment{thebibliography}[1]
	{\begin{list}{\arabic{enumi}.}
	{\usecounter{enumi}\setlength{\parsep}{0pt}
\setlength{\leftmargin 1.25cm}{\rightmargin 0pt}
	 \setlength{\itemsep}{0pt} \settowidth
	{\labelwidth}{#1.}\sloppy}}{\end{list}}

\topsep=0in\parsep=0in\itemsep=0in
\parindent=1.5pc

\newcounter{itemlistc}
\newcounter{romanlistc}
\newcounter{alphlistc}
\newcounter{arabiclistc}
\newenvironment{itemlist}
    	{\setcounter{itemlistc}{0}
	 \begin{list}{$\bullet$}
	{\usecounter{itemlistc}
	 \setlength{\parsep}{0pt}
	 \setlength{\itemsep}{0pt}}}{\end{list}}

\newenvironment{romanlist}
	{\setcounter{romanlistc}{0}
	 \begin{list}{$($\roman{romanlistc}$)$}
	{\usecounter{romanlistc}
	 \setlength{\parsep}{0pt}
	 \setlength{\itemsep}{0pt}}}{\end{list}}

\newenvironment{alphlist}
	{\setcounter{alphlistc}{0}
	 \begin{list}{$($\alph{alphlistc}$)$}
	{\usecounter{alphlistc}
	 \setlength{\parsep}{0pt}
	 \setlength{\itemsep}{0pt}}}{\end{list}}

\newenvironment{arabiclist}
	{\setcounter{arabiclistc}{0}
	 \begin{list}{\arabic{arabiclistc}}
	{\usecounter{arabiclistc}
	 \setlength{\parsep}{0pt}
	 \setlength{\itemsep}{0pt}}}{\end{list}}

\newcommand{\fcaption}[1]{
        \refstepcounter{figure}
        \setbox\@tempboxa = \hbox{\tenrm Fig.~\thefigure. #1}
        \ifdim \wd\@tempboxa > 6in
           {\begin{center}
        \parbox{6in}{\tenrm\baselineskip=12pt Fig.~\thefigure. #1 }
            \end{center}}
        \else
             {\begin{center}
             {\tenrm Fig.~\thefigure. #1}
              \end{center}}
        \fi}

\newcommand{\tcaption}[1]{
        \refstepcounter{table}
        \setbox\@tempboxa = \hbox{\tenrm Table~\thetable. #1}
        \ifdim \wd\@tempboxa > 6in
           {\begin{center}
        \parbox{6in}{\tenrm\baselineskip=12pt Table~\thetable. #1 }
            \end{center}}
        \else
             {\begin{center}
             {\tenrm Table~\thetable. #1}
              \end{center}}
        \fi}

\def\@citex[#1]#2{\if@filesw\immediate\write\@auxout
	{\string\citation{#2}}\fi
\def\@citea{}\@cite{\@for\@citeb:=#2\do
	{\@citea\def\@citea{,}\@ifundefined
	{b@\@citeb}{{\bf ?}\@warning
	{Citation `\@citeb' on page \thepage \space undefined}}
	{\csname b@\@citeb\endcsname}}}{#1}}

\newif\if@cghi
\def\cite{\@cghitrue\@ifnextchar [{\@tempswatrue
	\@citex}{\@tempswafalse\@citex[]}}
\def\citelow{\@cghifalse\@ifnextchar [{\@tempswatrue
	\@citex}{\@tempswafalse\@citex[]}}
\def\@cite#1#2{{$\null^{#1}$\if@tempswa\typeout
	{IJCGA warning: optional citation argument
	ignored: `#2'} \fi}}
\newcommand{\citeup}{\cite}

\def\fnm#1{$^{\mbox{\scriptsize #1}}$}
\def\fnt#1#2{\footnotetext{\kern-.3em
	{$^{\mbox{\sevenrm #1}}$}{#2}}}

\font\twelvebf=cmbx10 scaled\magstep 1
\font\twelverm=cmr10 scaled\magstep 1
\font\twelveit=cmti10 scaled\magstep 1
\font\elevenbfit=cmbxti10 scaled\magstephalf
\font\elevenbf=cmbx10 scaled\magstephalf
\font\elevenrm=cmr10 scaled\magstephalf
\font\elevenit=cmti10 scaled\magstephalf
\font\bfit=cmbxti10
\font\tenbf=cmbx10
\font\tenrm=cmr10
\font\tenit=cmti10
\font\ninebf=cmbx9
\font\ninerm=cmr9
\font\nineit=cmti9
\font\eightbf=cmbx8
\font\eightrm=cmr8
\font\eightit=cmti8

       {\normalsize \hfill
       \begin{tabbing}
       \`\begin{tabular}{l}
         SLAC--PUB--6673  \\
         hep--th/9409121 \\
         September 1994 \\
         (T) \\
         \end{tabular}
       \end{tabbing} }\vspace{8mm}
\thispagestyle{empty}

\centerline{\twelvebf CREATION OF A SCALAR POTENTIAL IN 2D DILATON
GRAVITY\footnote{Work supported by a grant of the DAAD and
by the Department of Energy,
 contract DE--AC03--76SF00515.}}
\vspace{0.8cm}
\centerline{\tenrm Klaus Behrndt\footnote{E-Mail:
                           behrndt@ifh.de}}
\baselineskip=13pt
\centerline{\tenit Stanford Linear Accelerator Center}
\baselineskip=12pt
\centerline{\tenit Stanford University, Stanford, California 94309, USA}
\vspace{0.9cm}

\renewcommand{\arraystretch}{2.0}
\renewcommand{\thefootnote}{\alph{footnote}}
\newcommand{\be}{\begin{equation}}
\newcommand{\ee}{\end{equation}}
\newcommand{\ba}{\begin{array}}
\newcommand{\ea}{\end{array}}
\newcommand{\vsf}{\vspace{5mm}}
\newcommand{\NP}[3]{{\em Nucl. Phys.}{ \bf B#1#2#3}}
\newcommand{\PRD}[2]{{\em Phys. Rev.}{ \bf D#1#2}}
\newcommand{\MPLA}[1]{{\em Mod. Phys. Lett.}{ \bf A#1}}
\newcommand{\PL}[3]{{\em Phys. Lett.}{ \bf B#1#2#3}}
\newcommand{\marpar}{\marginpar[!!!]{!!!}}
\newcommand{\lab}[2]{\label{#1#2}   (#1#2) \hfill }
\input epsf.tex
\vspace{10mm}

\abstracts{
We investigate quantum corrections of the 2-d dilaton
gravity near the singularity. Our motivation comes from
a s-wave reduced cosmological solution which is classically
singular in the scalar fields (dilaton and moduli). As result we find,
that the singularity disappears and a dilaton/moduli potential is
created.
}
\vspace*{10mm}

\vfill
\baselineskip=14pt

\begin{center}
Invited talk presented at \\
{\em 7th Marcel Grossmann Meeting} \\
{\em on General Relativity (MG 7) } \\
Stanford, CA, July 24  -- 30 , 1994
\end{center}

\vfill

\newpage

\twelverm   
\baselineskip=14pt

In two dimensions (2-D) the dilaton gravity could be formulated as a
quantum theory in the last years. This opens the possibility to
quantize higher dimensional theories near regions where a 2-D part
factorizes. As a first step one can quantize only the 2-D part and
leave the dynamical fields living in the other dimensions as a
classical background. An example is the s-wave reduction of higher
dimensional theories with a spherical symmetry, e.g.\ coming from
black holes or cosmological solutions. In this talk we are not going
to describe the black hole background. Instead, our interest in this
investigation comes from a special cosmological
solution\cite{clw,be/fo}, that can be obtained by a dimensional
reduction. This solution has singularities in the scalar fields
(dilaton and modulus) and our aim is to discuss quantum corrections
near the singularity (details can be found in Ref.\ 4). A crucial
property of this solution is that near the singularity it factorizes
in a smooth 3-D spherical part and a divergent 2-D part. If we perform
a s-wave reduction we find that the divergency is controlled by the
known 2-D dilaton gravity
\be \label{12}
S^{(2)} = \int d^2z \sqrt{g} e^{-2 \phi} \left( R^{(2)}
+ 4 (\partial  \phi)^2 + \lambda \right)
\ee
where $\phi$ is the 2-D dilaton and $\lambda$ is constant.
The classical solution in conformal coordinates is given
by
\be  \label{13}
ds^2 = e^{2 \sigma} dz^+ dz^-  \qquad , \qquad e^{-2 \phi} \sim e^{-2 \sigma}
= u - \lambda z^+ z^- \qquad (u=const.)
\ee
The singularity of this solution is in the strong coupling region
($\phi\rightarrow + \infty$) whereas in weak coupling region
($\phi\rightarrow - \infty$) it behaves smooth.

Before we turn to the discussion of a scalar potential let us shortly
summarize the quantization procedure
(we are now following here the notation of
de Alwis\cite{deal}). Choosing the conformal gauge: $g_{ab} = e^{2
\sigma} \hat{g}_{ab}$ and performing the field redefinition
\be  \label{22}
x = \frac{1}{\sqrt{4 \kappa}}\left( - \sqrt{\kappa^2 + 4 e^{-4 \phi}}
  + \sqrt{\kappa} \mbox{arcsinh} \frac{\kappa}{2} e^{2\phi} \right) \ ,\
y = \sqrt{\kappa} \left( \sigma - \frac{1}{\kappa} e^{-2 \phi}
  - \phi \right)
\ee
we can write the 2d model as (including terms from the functional
integration measure)
\be \label{18}
S^{(2)} = \int d^2 z \sqrt{\hat{g}} \left[ (\partial x)^2 - (\partial y)^2
 + \hat{R} \, \Phi(x,y) + T(x,y) \right] \ .
\ee
However, the function $\Phi(x,y)$ and $T(x,y)$ are not arbitrary.
The requirement of independence of the reference metric $\hat{g}_{ab}$
has the consequence that the 2-D theory has to be conformally invariant.
The simplest choice is to take a linear dilaton $\Phi$ and a exponential
tachyon $T$
\be
\Phi = \sqrt{\kappa} y \qquad , \qquad  T = \lambda \, e^{\frac{2}{
\sqrt{\kappa}} (x-y)}\ .
\ee
With this choice we have a well defined
2-D quantum theory (mathematically the same as the non-critical string
theory in one dimension). Now, one defines the quantum theory in terms
of these $x$ and $y$ variables and regards Eq.~(\ref{12}) as the
classical limit. As solution of the equation of motion in $x$ and $y$
one finds (if we restrict ourselves on solutions depending on the
product $z^+ z^-$ only)
\be  \label{21}
x = y = \frac{1}{\sqrt{\kappa}} \left(u - \lambda z^+ z^- \right)
\ee
($u=const.$). Using the transformation (\ref{22}) we can express this
solution by $\phi$ and $\sigma$. In the weak coupling limit ($e^{2
\phi} \ll 1$) we have the desired classical solution (\ref{13}).
But our original singularity appeared in the strong coupling region.
In this limit ($e^{2 \phi} \gg 1$) we obtain
$
\phi = - \frac{1}{\kappa} ( u - \lambda z^+ z^-) $, $ \sigma =
  \frac{1}{\kappa} e^{-2 \phi}
$.
Thus, after incorporation of quantum corrections ($\sim
{\cal{O}}(e^{2 \phi})$) the solution becomes smooth also in the
strong coupling region\cite{deal}.

One can now ask, whether quantum corrections can form a potential in
the scalar fields $x$ and $y$ (or $\phi$ and $\sigma$ resp.).  A
potential in our original action (\ref{12}) corresponds to an
additional tachyon contribution. The tachyon we have discussed so far
is only {\em one} possibility. The most general tachyon field is a
combination of the solutions of the Weyl invariance condition, that
are given by
\be  \label{25}
\ba{ll}
\Phi(x,y) = a x + b y & \qquad \mbox{with} \qquad a^2 - b^2 =
 - \kappa \qquad ,\\
T(x,y) \sim e^{\alpha x + \beta y}  & \qquad \mbox{with} \qquad
    \frac{1}{2} (\alpha^2  - \beta^2) -  a \alpha +
   b \beta - 2 = 0 \ .
\ea
\ee
In order to get the right classical limit we set furthermore
$a=0$. But there is also another parameterization for the
tachyon\cite{kawa}. Using the mass shell condition we can replace $\alpha$
or $\beta$ and then we can expand the tachyon field in powers of the
remaining $\alpha$ or $\beta$.
After this procedure we find an infinite set of tachyon fields which are
parameterized by two integers $m$ and $n$. Because the corresponding
tachyon equation is linear every term of this expansion fulfills the
equation, too. If we restrict ourselves on
$\kappa = \frac{24-N}{6} = 4$ (i.e.~$N=0$) these additional terms are
\be \label{26}
T_{2}^{(n)} = (y-x)^n \, e^{2 x}  \quad , \quad T_{3}^{(m)} = (x\pm y)^m
 \, e^{2 y} \ .
\ee
Instead of Eq.~(\ref{25}) we have now as general tachyon field $T(x,y)$
\be  \label{27}
T(x,y) = \lambda e^{\frac{2}{\sqrt{\kappa}} (x-y)} + \sum_{(n,m)}
(\mu_{2}^{n} T_{2}^{(n)} + \mu_{3}^{m} T_{3}^{(m)} )
\ee
where the function $x$ and $y$ are given by the Eq.~(\ref{22}) (the
term $T_{2}^{(0)}$ was already discussed in Refs.~1 and 6).
A remarkable property of these terms is, that they have
in the classical limit ($\phi \rightarrow - \infty$) the typical
non-perturbative structure: $T_{2,3} \sim e^{-\frac{1}{2} e^{-2\phi}} \sim
e^{- \frac{1}{(2 g^2_s)}} $,
where $g_s = e^{\phi}$ is the string coupling constant. On the other side, in
the strong coupling region ($\phi \rightarrow \infty$) we have:
$
T_{2} \sim e^{4 \phi} \rightarrow \infty $ , $
T_{3} \sim e^{-2 \phi} \rightarrow 0  $.
Therefore, these terms vanish very rapidly in the weak coupling
(classical) region and become important in the strong coupling
region. Furthermore, since $x$ and $y$ are functions of the scalar
fields these terms represent a potential in $\phi$ and $\sigma$.  So,
the quantized theory (\ref{18}) differs from the classical theory
(\ref{12}) not only by a modification of the kinetic terms of $\phi$
and $\sigma$ but also by an additional potential in the scalar
fields. What does this mean for a cosmological solution?  There, the
scalar fields $\phi$ and $\sigma$ correspond to a dilaton field and a
modulus field. As we have pointed out the cosmological solution is
classically singular in the scalar fields. This semi-classical
quantization (we have only quantized the scalar fields) showed that
the singularity disappears and, in addition, a dilaton/moduli
potential is created.  It remains an open question whether this
potential can yield sufficient inflation.  In order to discuss this
question one has to transform the theory back to the 4-D Einstein
frame and has to show that the resulting potential has a flat
direction which in turn give an extended
inflation\cite{ga/qu}. Probably, this is possible for a suitable
choice of the constants $\mu_{2,3}^{m,n}$. However, normally in
discussing of non-perturbative corrections one imposes further string
symmetries to restrict the possible contributions and it deserves
further investigations to show that this will not destroy a flat
direction.

\end{document}